\documentclass[conference]{IEEEtran}
\ifCLASSINFOpdf
\else
   \usepackage[dvips]{graphicx}
   \newcommand{\bysame}{%
    \leavevmode\hbox to 3em{\hrulefill}\,}
\fi
\begin{document}
%
\title{Reduction of Error-Trellises for Tail-Biting Convolutional Codes Using Shifted Error-Subsequences}

\author{\IEEEauthorblockN{Masato Tajima}
\IEEEauthorblockA{Graduate School of Sci. and Eng.\\
University of Toyama\\
3190 Gofuku, Toyama 930-8555, Japan\\
Email: tajima@eng.u-toyama.ac.jp}
\and
\IEEEauthorblockN{Koji Okino}
\IEEEauthorblockA{Information Technology Center\\
University of Toyama\\
3190 Gofuku, Toyama 930-8555, Japan\\
Email: okino@itc.u-toyama.ac.jp}
\and
\IEEEauthorblockN{Tatsuto Murayama}
\IEEEauthorblockA{Graduate School of Sci. and Eng.\\
University of Toyama\\
3190 Gofuku, Toyama 930-8555, Japan\\
Email: murayama@eng.u-toyama.ac.jp}
}

%


\maketitle

\begin{abstract}
In this paper, we discuss the reduction of error-trellises for tail-biting convolutional codes. In the case where some column of a parity-check matrix has a monomial factor $D^l$, we show that the associated tail-biting error-trellis can be reduced by cyclically shifting the corresponding error-subsequence by $l$ (i.e., the power of $D$) time units. We see that the resulting reduced error-trellis is again tail-biting. Moreover, we show that reduction is also possible using backward-shifted error-subsequences.
\end{abstract}


%
\IEEEpeerreviewmaketitle

\section{Introduction}
Tail-biting is a technique by which a convolutional code can be used to construct a block code without any loss of rate [4], [6], [14]. Let $C_{tb}$ be a tail-biting convolutional code with an $N$-section code-trellis $T_{tb}^{(c)}$. The fundamental idea behind tail-biting is that the encoder starts and ends in the same state, i.e., $\mbox{\boldmath $\beta$}_0=\mbox{\boldmath $\beta$}_N$ ($\mbox{\boldmath $\beta$}_k$ is the encoder state at time $k$). Suppose that $T_{tb}^{(c)}$ has $\Sigma_0$ initial (or final) states, then it is composed of $\Sigma_0$ subtrellises, each having the same initial and final states. We call these subtrellises tail-biting code subtrellises. For example, a tail-biting code-trellis of length $N=4$ based on the generator matrix
\begin{equation}
G_1(D)=(D+D^2, D^2, 1+D)
\end{equation}
is shown in Fig.1. Since $\Sigma_0=4$, this tail-biting code-trellis is composed of $4$ code subtrellises. In Fig.1, bold lines correspond to the code subtrellis with $\mbox{\boldmath $\beta$}_0=\mbox{\boldmath $\beta$}_4=(1,1)$.
\begin{figure}[tb]
\begin{center}
\includegraphics[width=8.0cm,clip]{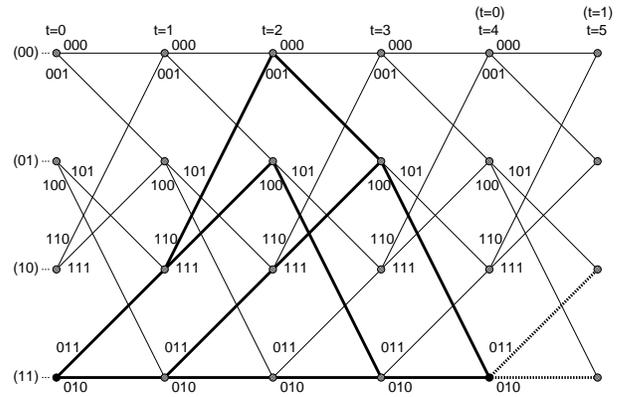}
\end{center}
\caption{Tail-biting code-trellis based on $G_1(D)$.}
\label{Fig.1}
\end{figure}
On the other hand, it is reasonable to think that an error-trellis $T_{tb}^{(e)}$ for the tail-biting convolutional code $C_{tb}$ can equally be constructed. In this case, each error subtrellis should have the same initial and final states like a code subtrellis. In our previous works [11], [12], taking this property into consideration, we have presented an error-trellis construction for tail-biting convolutional codes. For example, consider the above case. The parity-check matrix $H_1(D)$ associated with $G_1(D)$ is given by
\begin{equation}
H_1(D)=\left(
\begin{array}{ccc}
1& 0 & D \\
D& 1+D & 0 
\end{array}
\right) .
\end{equation}
Let
$$
\mbox{\boldmath $z$}=\mbox{\boldmath $z$}_1~\mbox{\boldmath $z$}_2~\mbox{\boldmath $z$}_3~\mbox{\boldmath $z$}_4=110~101~101~011
$$
be the received data. In this case, using the method in [11], the tail-biting error-trellis corresponding to the code-trellis in Fig.1 can be constructed as is shown in Fig.2, where bold lines correspond to the error subtrellis with $\mbox{\boldmath $\sigma$}_0=\mbox{\boldmath $\sigma$}_4=(1,0)$.
\begin{figure}[tb]
\begin{center}
\includegraphics[width=8.0cm,clip]{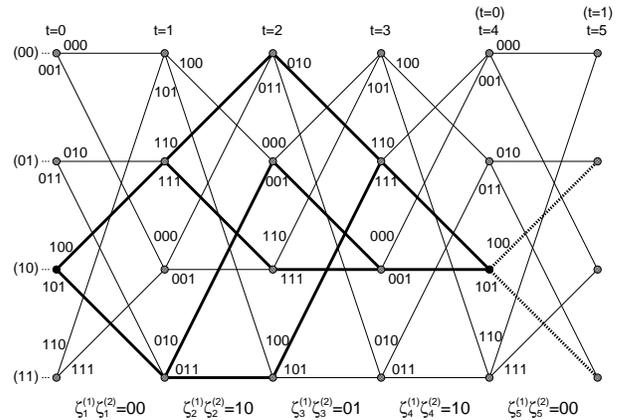}
\end{center}
\caption{Tail-biting error-trellis based on $H_1^T(D)$.}
\label{Fig.2}
\end{figure}
\par
On the other hand [9], note that the third column of $H_1(D)$ has the monomial factor $D$. Let $\mbox{\boldmath $e$}_k=(e_k^{(1)}, e_k^{(2)}, e_k^{(3)})$ and $\mbox{\boldmath $\zeta$}_k=(\zeta_k^{(1)}, \zeta_k^{(2)})$ be the time-$k$ error and syndrome, respectively. We have the following modification ($T$ means transpose):
\begin{eqnarray}
\mbox{\boldmath $\zeta$}_k &=&\mbox{\boldmath $e$}_kH_1^T(D) \\
&=& (e_k^{(1)}, e_k^{(2)}, e_k^{(3)})\left(
\begin{array}{ccc}
1& D \\
0& 1+D \\
D& 0
\end{array}
\right) \nonumber \\
&=& (e_k^{(1)}, e_k^{(2)}, De_k^{(3)})\left(
\begin{array}{ccc}
1& D \\
0& 1+D \\
1& 0
\end{array}
\right) \nonumber \\
&\stackrel{\triangle}{=}& \mbox{\boldmath $\tilde e$}_k\tilde H_1^{T}(D) ,
\end{eqnarray}
where $\mbox{\boldmath $\tilde e$}_k=(e_k^{(1)}, e_k^{(2)}, \tilde e_k^{(3)})$ and $\tilde e_k^{(3)}$ is defined as $De_k^{(3)}=e_{k-1}^{(3)}$. Since the overall constraint length $\tilde \nu^{\perp}$ of
\begin{equation}
\tilde H_1(D)=\left(
\begin{array}{ccc}
1& 0 & 1  \\
D& 1+D & 0 
\end{array}
\right)
\end{equation}
is one, the above equation implies that the tail-biting error-trellis in Fig.2 can be reduced by shifting the subsequence $\{e_k^{(3)}\}$ by the unit time.
\par
In this paper, taking the above example into account, we discuss the reduction of error-trellises for tail-biting convolutional codes. It is assumed that some ($j$th) column of a parity-check matrix $H(D)$ has a monomial factor $D^{l_j}$. In this case, we show that the associated tail-biting error-trellis can be reduced by cyclically shifting the $j$th component $e_k^{(j)}$ by $l_j$ time units. We also show that the resulting reduced error-trellis is again tail-biting. We see that a kind of ``periodicity'' inherent in tail-biting trellises plays a key role in our discussion.

\section{Preliminaries}
In this paper, we always assume that the underlying field is $F=\mbox{GF}(2)$. Let $G(D)$ be a generator matrix of an $(n, n-m)$ convolutional code $C$. Let $H(D)$ be a corresponding $m \times n$ parity-check matrix of $C$. Both $G(D)$ and $H(D)$ are assumed to be canonical [1], [5]. Denote by $\nu^{\perp}$ the overall constraint length of $H(D)$ and by $M$ the memory length of $H(D)$ (i.e., the maximum degree among the polynomials of $H(D)$). Then $H(D)$ is expressed as
\begin{equation}
H(D)=H_0+H_1D+ \cdots +H_MD^M .
\end{equation}

\subsection{Adjoint-Obvious Realization of a Syndrome Former}
Consider the adjoint-obvious realization (observer canonical form [2], [3]) of the syndrome former $H^T(D)$. Let $\mbox{\boldmath $e$}_k=(e_k^{(1)}, e_k^{(2)},\cdots, e_k^{(n)})$ and $\mbox{\boldmath $\zeta$}_k=(\zeta_k^{(1)}, \zeta_k^{(2)}, \cdots, \zeta_k^{(m)})$ be the input error and the corresponding output syndrome at time $k$, respectively. Denote by $\sigma_{kp}^{(q)}$ the contents of the memory elements in the above realization. (If a memory element is missing, the corresponding $\sigma_{kp}^{(q)}$ is set to zero.) Using $\sigma_{kp}^{(q)}$, the syndrome-former state at time $k$ is defined as
\begin{equation}
\mbox{\boldmath $\sigma$}_k\stackrel{\triangle}{=}(\sigma_{k1}^{(1)}, \cdots, \sigma_{k1}^{(m)}, \cdots, \sigma_{kM}^{(1)}, \cdots, \sigma_{kM}^{(m)}) .
\end{equation}
({\it Remark:} The effective size of $\mbox{\boldmath $\sigma$}_k$ is equal to $\nu^{\perp}$.)
\par
For example, Fig.3 illustrates the adjoint-obvious realization of the syndrome former $H_2^T(D)$ [1], where
\begin{equation}
H_2(D)=\left(
\begin{array}{ccc}
D^2+D^3& D & 1 \\
D^2& 1+D+D^2 & D^2 
\end{array}
\right) .
\end{equation}
Hence, we have
\begin{equation}
\mbox{\boldmath $\sigma$}_k=(\sigma_{k1}^{(1)}, \sigma_{k1}^{(2)}, \sigma_{k2}^{(1)}, \sigma_{k2}^{(2)}, \sigma_{k3}^{(1)}, 0) .
\end{equation}
Note that the effective size of $\mbox{\boldmath $\sigma$}_k$ is $\nu^{\perp}=5$.
\begin{figure}[tb]
\begin{center}
\includegraphics[width=7.5cm,clip]{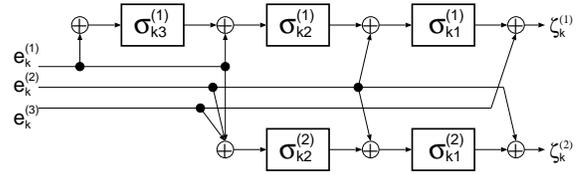}
\end{center}
\caption{Adjoint-obvious realization (observer canonical form) of syndrome former $H_2^T(D)$.}
\label{Fig.3}
\end{figure}
\par
Under the above conditions [7], [8], we have
\begin{eqnarray}
\mbox{\boldmath $\sigma$}_k &=& (\mbox{\boldmath $\sigma$}_k^{(1)}, \mbox{\boldmath $\sigma$}_k^{(2)}, \cdots, \mbox{\boldmath $\sigma$}_k^{(M)}) \nonumber \\
&=& (\mbox{\boldmath $e$}_{k-M+1}, \cdots, \mbox{\boldmath $e$}_{k-1}, \mbox{\boldmath $e$}_k) \nonumber \\
&& \times \left(
\begin{array}{cccc}
H_M^T & \scriptstyle{\ldots} & 0 & 0 \\
\scriptstyle{\ldots} & \scriptstyle{\ldots} & \scriptstyle{\ldots} & \scriptstyle{\ldots} \\
H_2^T & \scriptstyle{\ldots} & H_M^T & 0 \\
H_1^T & \scriptstyle{\ldots} & H_{M-1}^T & H_M^T
\end{array}
\right) .
\end{eqnarray}
Note that $\mbox{\boldmath $\sigma$}_k$ has an alternative expression:
\begin{equation}
\mbox{\boldmath $\sigma$}_k=(\mbox{\boldmath $\sigma$}_{k-1}^{(2)}, \cdots, \mbox{\boldmath $\sigma$}_{k-1}^{(M)}, \mbox{\boldmath $0$})+\mbox{\boldmath $e$}_k(H_1^T, H_2^T, \cdots, H_M^T) .
\end{equation}
Similarly, $\mbox{\boldmath $\zeta$}_k$ is expressed as
\begin{eqnarray}
\mbox{\boldmath $\zeta$}_k &=& \mbox{\boldmath $e$}_{k-M}H_M^T+\cdots+\mbox{\boldmath $e$}_{k-1}H_1^T+\mbox{\boldmath $e$}_kH_0^T \\
&=& \mbox{\boldmath $\sigma$}_{k-1}^{(1)}+\mbox{\boldmath $e$}_kH_0^T .
\end{eqnarray}

\subsection{Dual States}
The encoder states can be labeled by the syndrome-former states (i.e., dual states [2]). The dual state $\mbox{\boldmath $\beta$}_k^*$ corresponding to an encoder state $\mbox{\boldmath $\beta$}_k$ is obtained by replacing $\mbox{\boldmath $e$}_k$ in $\mbox{\boldmath $\sigma$}_k$ by $\mbox{\boldmath $y$}_k=\mbox{\boldmath $u$}_kG(D)$ ($\mbox{\boldmath $y$}_k$ is the code symbol at time $k$ corresponding to the information symbol $\mbox{\boldmath $u$}_k$). We have
\begin{eqnarray}
\mbox{\boldmath $\beta$}_k^* &=& (\mbox{\boldmath $y$}_{k-M+1}, \cdots, \mbox{\boldmath $y$}_{k-1}, \mbox{\boldmath $y$}_k) \nonumber \\
&& \times \left(
\begin{array}{cccc}
H_M^T & \scriptstyle{\ldots} & 0 & 0 \\
\scriptstyle{\ldots} & \scriptstyle{\ldots} & \scriptstyle{\ldots} & \scriptstyle{\ldots} \\
H_2^T & \scriptstyle{\ldots} & H_M^T & 0 \\
H_1^T & \scriptstyle{\ldots} & H_{M-1}^T & H_M^T
\end{array}
\right) .
\end{eqnarray}
\par
{\it Example 1:} Consider the parity-check matrix $H_1(D)$. We have
\begin{eqnarray}
H_1(D) &=& \left(
\begin{array}{ccc}
1& 0 & D \\
D & 1+D & 0 
\end{array}
\right) \nonumber \\
&=& \left(
\begin{array}{ccc}
1& 0 & 0 \\
0& 1 & 0 
\end{array}
\right)+\left(
\begin{array}{ccc}
0& 0 & 1 \\
1& 1 & 0 
\end{array}
\right)D \nonumber \\
&\stackrel{\triangle}{=}& H_0+H_1D .
\end{eqnarray}
Then the dual state corresponding to an encoder state $\mbox{\boldmath $\beta$}_k=(u_{k-1}, u_k)$ is obtained as follows.
\begin {eqnarray}
\mbox{\boldmath $\beta$}_k^* &=& \mbox{\boldmath $y$}_kH_1^T \nonumber \\
&=& (y_k^{(1)}, y_k^{(2)}, y_k^{(3)}) \left(
\begin{array}{cc}
0 & 1 \\
0 & 1 \\
1 & 0 
\end{array}
\right) \nonumber \\
&=& (y_k^{(3)}, y_k^{(1)}+y_k^{(2)}) \nonumber \\
&=& (u_{k-1}+u_k, u_{k-1}) .
\end{eqnarray}

\subsection{Error-Trellises for Tail-Biting Convolutional Codes}
Suppose that a tail-biting code-trellis based on $G(D)$ is defined in $[0, N]$, where $N\geq M$. In this case, the corresponding tail-biting error-trellis based on $H^T(D)$ is constructed as follows [11].
\par
{\it Step 1:} Let $\mbox{\boldmath $z$}=\{\mbox{\boldmath $z$}_k\}_{k=1}^N$ be a received data. Denote by $\mbox{\boldmath $\sigma$}_0$ the initial state of the syndrome former $H^T(D)$. Let $\mbox{\boldmath $\sigma$}_{fin}(=\mbox{\boldmath $\sigma$}_N)$ be the final syndrome-former state corresponding to the input {\boldmath $z$}. Note that $\mbox{\boldmath $\sigma$}_{fin}$ is independent of $\mbox{\boldmath $\sigma$}_0$ and is uniquely determined only by {\boldmath $z$}.
\par
{\it Step 2:} Set $\mbox{\boldmath $\sigma$}_0$ (i.e., the initial state of the syndrome former) to $\mbox{\boldmath $\sigma$}_{fin}$ and input {\boldmath $z$} to the syndrome former again. Here, suppose that the syndrome sequence $\mbox{\boldmath $\zeta$}=\{\mbox{\boldmath $\zeta$}_k\}_{k=1}^N$ is obtained. ({\it Remark:} $\mbox{\boldmath $\zeta$}_k~(k\geq M+1)$ has been obtained in Step 1.)
\par
{\it Step 3:} Concatenate the error-trellis modules corresponding to the syndromes $\mbox{\boldmath $\zeta$}_k$. Then we have the tail-biting error-trellis.
\par
{\it Example 2:} Again, consider the parity-check matrix $H_1(D)$. Let
\begin{equation}
\mbox{\boldmath $z$}=\mbox{\boldmath $z$}_1~\mbox{\boldmath $z$}_2~\mbox{\boldmath $z$}_3~\mbox{\boldmath $z$}_4=110~101~101~011
\end{equation}
be the received data. According to Step 1, let us input {\boldmath $z$} to the syndrome former $H_1^T(D)$. Then we have $\mbox{\boldmath $\sigma$}_{fin}=(1,1)$. Next, set $\mbox{\boldmath $\sigma$}_0$ to $\mbox{\boldmath $\sigma$}_{fin}=(1,1)$ and input {\boldmath $z$} to the syndrome former again. In this case, the syndrome sequence
\begin{equation}
\mbox{\boldmath $\zeta$}=\mbox{\boldmath $\zeta$}_1~\mbox{\boldmath $\zeta$}_2~\mbox{\boldmath $\zeta$}_3~\mbox{\boldmath $\zeta$}_4=00~10~01~10
\end{equation}
is obtained. The tail-biting error-trellis is constructed by concatenating the error-trellis modules corresponding to $\mbox{\boldmath $\zeta$}_k$. The resulting tail-biting error-trellis is shown in Fig.2.
\par
With respect to the correspondence between tail-biting code subtrellises and tail-biting error subtrellises, we have the following [11], [12].
\newtheorem{pro}{Proposition}
\begin{pro}
Let $\mbox{\boldmath $\beta$}_0(=\mbox{\boldmath $\beta$}_N)=\mbox{\boldmath $\beta$}$ be the initial (final) state of a tail-biting code subtrellis. Then the initial (final) state of the corresponding tail-biting error subtrellis is given by $\mbox{\boldmath $\sigma$}_{fin}+\mbox{\boldmath $\beta$}^*$.
\end{pro}
\par
{\it Example 2 (Continued):} Consider the tail-biting error-trellis in Fig.2. In this example, we have $\mbox{\boldmath $\sigma$}_{fin}=(1,1)$. The corresponding tail-biting code-trellis based on $G_1(D)$ is shown in Fig.1. In Fig.1, take notice of the code subtrellis with initial (final) state $\mbox{\boldmath $\beta$}=(1,1)$ (bold lines). The dual state of $\mbox{\boldmath $\beta$}=(1,1)$ is calculated as $\mbox{\boldmath $\beta$}^*=(u_{-1}+u_0, u_{-1})=(1+1,1)=(0,1)$. Hence, the initial (final) state of the corresponding error subtrellis is given by $\mbox{\boldmath $\sigma$}_{fin}+\mbox{\boldmath $\beta$}^*=(1, 1)+(0, 1)=(1, 0)$ (bold lines in Fig.2).

\section{Reduction of Tail-Biting Error-Trellises}
\subsection{Error-Trellis Reduction Using Shifted Error-Subsequences}
Consider the example in Section I. Noting the relation $\tilde e_k^{(3)}=e_{k-1}^{(3)}$, we cyclically shift the third component of each $\mbox{\boldmath $z$}_k$ to the right by the unit time. Then we have the modified received data
\begin{equation}
\mbox{\boldmath $\tilde z$}=\mbox{\boldmath $\tilde z$}_1~\mbox{\boldmath $\tilde z$}_2~\mbox{\boldmath $\tilde z$}_3~\mbox{\boldmath $\tilde z$}_4=111~100~101~011 .
\end{equation}
Applying the method in Section II-C, we can construct a reduced tail-biting error-trellis. According to Step 1, let us input {\boldmath $\tilde z$} to the syndrome former $\tilde H_1^T(D)$. Then we have $\mbox{\boldmath $\tilde \sigma$}_{fin}=(1)$. Next, set $\mbox{\boldmath $\tilde \sigma$}_0$ to $\mbox{\boldmath $\tilde \sigma$}_{fin}=(1)$ and input {\boldmath $\tilde z$} to the syndrome former again. In this case, the same syndrome sequence as the original one (i.e., $\mbox{\boldmath $\zeta$}=00~10~01~10$) is obtained. The reduced tail-biting error-trellis is constructed by concatenating the reduced error-trellis modules corresponding to $\mbox{\boldmath $\zeta$}_k$. The resulting tail-biting error-trellis is shown in Fig.4. Here let us examine how a tail-biting error subtrellis is embedded in the corresponding reduced error-trellis. For the purpose, take notice of the subtrellis with initial (final) state $(1, 0)$ (bold lines in Fig.2). First, consider where the state $(1, 0)$ is mapped to. In the original error-trellis, the final state $\mbox{\boldmath $\sigma$}_N$ is expressed as
\begin {equation}
\mbox{\boldmath $\sigma$}_N=\mbox{\boldmath $e$}_NH_1^T=(e_N^{(3)}, e_N^{(1)}+e_N^{(2)}) .
\end{equation}
Using the relation $e_N^{(3)}=\tilde e_{N+1}^{(3)}$, $\mbox{\boldmath $\sigma$}_N=(e_N^{(3)}, e_N^{(1)}+e_N^{(2)})$ is modified as $(\tilde e_{N+1}^{(3)}, e_N^{(1)}+e_N^{(2)})$. Since the subscript $N+1~(>N)$ is inappropriate for the state at time $N$, we have
\begin {equation}
\mbox{\boldmath $\tilde \sigma$}_N=e_N^{(1)}+e_N^{(2)}~(=\mbox{\boldmath $\tilde e$}_N\tilde H_1^T) .
\end{equation}
({\it Remark:} We have $\mbox{\boldmath $\tilde e$}_N\tilde H_1^T=(0, e_N^{(1)}+e_N^{(2)})$. Hence, the first component can be deleted.) That is, state $\mbox{\boldmath $\sigma$}_4=(1, 0)$ is mapped to $\mbox{\boldmath $\tilde \sigma$}_4=(0)$.
\par
Next, consider an arbitrary error-path
$$
\mbox{\boldmath $e$}_p=\mbox{\boldmath $e$}_1~\mbox{\boldmath $e$}_2~\mbox{\boldmath $e$}_3~\mbox{\boldmath $e$}_4
$$
in the subtrellis with initial (final) state $(1, 0)$. Here take notice of two sections from $t=0$ to $t=1$ and from $t=3$ to $t=4$. Note that these are adjacent sections in the circular error-trellis. From $\mbox{\boldmath $\sigma$}_4=(e_4^{(3)}, e_4^{(1)}+e_4^{(2)})=(1, 0)$, we have $e_4^{(3)}=1$. Since the third component of each $\mbox{\boldmath $e$}_k$ is cyclically shifted to the right by the unit time, $e_1^{(3)}$ is replaced by $e_4^{(3)}=1$. That is, the third label on the first branch of the error-path in the reduced trellis must be $1$. By taking account of these conditions, we have four admissible error-paths:
\begin{eqnarray}
\mbox{\boldmath $\tilde e$}_{p_1} &=& 101~110~010~110 \nonumber \\
\mbox{\boldmath $\tilde e$}_{p_2} &=& 101~110~111~001 \nonumber \\
\mbox{\boldmath $\tilde e$}_{p_3} &=& 101~011~000~001 \nonumber \\
\mbox{\boldmath $\tilde e$}_{p_4} &=& 101~011~101~110 . \nonumber
\end{eqnarray}
These paths are denoted with bold lines in Fig.4. Since a planar trellis is used in Fig.4, the first segment (i.e., $101$)
 of each error-path is added to it as a tail. If a circular trellis is used, this augmentation is unnecessary. In this way, we obtain the reduced tail-biting error subtrellis. The original error-paths are restored by noting the relation $e_k^{(3)}=\tilde e_{k+1}^{(3)}$. That is, we only need to cyclically shift the third component of each $\mbox{\boldmath $\tilde e$}_k=(e_k^{(1)}, e_k^{(2)}, \tilde e_k^{(3)})$ to the left by the unit time. As a result, four error-paths
\begin{eqnarray}
\mbox{\boldmath $e$}_{q_1} &=& 100~110~010~111 \nonumber \\
\mbox{\boldmath $e$}_{q_2} &=& 100~111~111~001 \nonumber \\
\mbox{\boldmath $e$}_{q_3} &=& 101~010~001~001 \nonumber \\
\mbox{\boldmath $e$}_{q_4} &=& 101~011~100~111 \nonumber
\end{eqnarray}
are obtained. We see that these paths completely coincide with those in Fig.2.
\begin{figure}[tb]
\begin{center}
\includegraphics[width=8.0cm,clip]{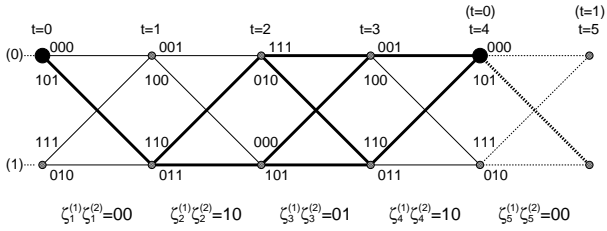}
\end{center}
\caption{Reduced tail-biting error-trellis based on $\tilde H_1^T(D)$.}
\label{Fig.4}
\end{figure}

\subsection{General Cases}
The argument in the previous subsection, though it was presented in terms of a specific example, is entirely general. Hence, it can be directly extended to general cases. Suppose that a specific ($j$th) column of $H(D)$ has the form
\begin{equation}
\left(
\begin{array}{cccc}
D^{l_j}\tilde h_{1j}(D) &D^{l_j}\tilde h_{2j}(D) &\ldots &D^{l_j}\tilde h_{mj}(D)
\end{array}
\right)^T ,
\end{equation}
where $1\leq l_j \leq M$. ({\it Remark:} A more general case where each column has the above form can also be treated.) Let $\tilde H(D)$ be the modified version of $H(D)$ with the $j$th column being replaced by
\begin{equation}
\left(
\begin{array}{cccc}
\tilde h_{1j}(D) &\tilde h_{2j}(D) &\ldots &\tilde h_{mj}(D)
\end{array}
\right)^T .
\end{equation}
$\tilde H(D)$ is assumed to be canonical. In this case, the reduction of a tail-biting error-trellis is accomplished as follows.
\par
{\it (i) Fundamental relation:} Denote by $\mbox{\boldmath $e$}_k=(e_k^{(1)}, \cdots, e_k^{(n)})$ and $\mbox{\boldmath $\zeta$}_k=(\zeta_k^{(1)}, \cdots, \zeta_k^{(m)})$ the time-$k$ error and syndrome, respectively. Also, let $\mbox{\boldmath $\tilde e$}_k\stackrel{\triangle}{=}(e_k^{(1)}, \cdots, e_{<k-l_j>}^{(j)}, \cdots, e_k^{(n)})$, where $<t>$ denotes $t\:\mbox{mod}\:N$. Then we have
\begin{equation}
\mbox{\boldmath $\zeta$}_k=\mbox{\boldmath $\tilde e$}_k\tilde H^T(D) .
\end{equation}
\par
{\it (ii) Construction of reduced tail-biting error-trellises:} Let
\begin{equation}
\mbox{\boldmath $z$}=\{\mbox{\boldmath $z$}_k\}_{k=1}^N=\{(z_k^{(1)}, \cdots, z_k^{(j)}, \cdots, z_k^{(n)})\}_{k=1}^N
\end{equation}
be a received data. We construct the modified received data
\begin{equation}
\mbox{\boldmath $\tilde z$}=\{\mbox{\boldmath $\tilde z$}_k\}_{k=1}^N\stackrel{\triangle}{=}\{(z_k^{(1)}, \cdots, z_{<k-l_j>}^{(j)}, \cdots, z_k^{(n)})\}_{k=1}^N
\end{equation}
by cyclically shifting the $j$th component of each $\mbox{\boldmath $z$}_k$ to the right by $l_j$ time units. By applying the method in Section II-C to the modified syndrome former $\tilde H^T(D)$ and the modified received data $\mbox{\boldmath $\tilde z$}$, a reduced tail-biting error-trellis is constructed. Note that the same syndrome sequence $\{\mbox{\boldmath $\zeta$}_k\}$ as for the tail-biting error-trellis based on $H^T(D)$ is obtained.
\par
{\it (iii) Reduced tail-biting error subtrellises:} Let $ST_{tb}^{(e)}$ be a tail-biting error subtrellis with initial (final) state $\mbox{\boldmath $\sigma$}_N$. $\mbox{\boldmath $\sigma$}_N$ can be expressed using $\{\mbox{\boldmath $e$}_t\}_{t=N-M+1}^N$ (cf. (10)). Here replace each $e_t^{(j)}~(N-M+1\leq t \leq N)$ by $\tilde e_{t+l_j}^{(j)}$ and delete those terms $\tilde e_t^{(j)}$ with subscript $t$ greater than $N$. Denote by $\mbox{\boldmath $\tilde \sigma$}_N$ the resulting state expression. In this case, state $\mbox{\boldmath $\sigma$}_N$ is mapped to state $\mbox{\boldmath $\tilde \sigma$}_N$ in the reduced tail-biting error-trellis.
\par
Consider the two trellis-sections from $t=0$ to $t=l_j$ and from $t=N-l_j$ to $t=N$. Note that these form a continuous section of length $2l_j$ in the circular error-trellis. Now we can solve Eq.(10) ($k=N$) given $\mbox{\boldmath $\sigma$}_N$. ({\it Remark:} $\{e_t^{(j)}\}_{t=N-l_j+1}^N$ is uniquely determined under a moderate condition on $H(D)$.) Since the $j$th component of each $\mbox{\boldmath $e$}_k$ is cyclically shifted to the right by $l_j$ time units, $e_t^{(j)}~(1\leq t \leq l_j)$ is replaced by $e_{N-l_j+t}^{(j)}$. That is, the $j$th component $\tilde e_t^{(j)}$ of the reduced path-segment $\mbox{\boldmath $\tilde e$}_t~(1\leq t \leq l_j)$ must be $e_{N-l_j+t}^{(j)}$. We call these segments ``admissible''. Then $ST_{tb}^{(e)}$ is embedded in the reduced tail-biting error subtrellis with initial (final) state $\mbox{\boldmath $\tilde \sigma$}_N$, where the path-segments in the first $l_j$ sections are restricted to admissible ones.
\par
{\it (iv) Restoration of the original error-paths:} The original error-paths are restored by noting the relation $e_k^{(j)}=\tilde e_{<k+l_j>}^{(j)}$. That is, for an error-path
\begin{equation}
\mbox{\boldmath $\tilde e$}=\{\mbox{\boldmath $\tilde e$}_k\}_{k=1}^N=\{(e_k^{(1)}, \cdots, \tilde e_k^{(j)}, \cdots, e_k^{(n)})\}_{k=1}^N ,
\end{equation}
we only need to cyclically shift the $j$th component of each $\mbox{\boldmath $\tilde e$}_k$ to the left by $l_j$ time units.
\par
We remark that {\boldmath $z$} has been periodically extended in both directions and this periodicity is fully used for tail-biting error-trellis construction. Now the relation $\mbox{\boldmath $\zeta$}_k=\mbox{\boldmath $e$}_kH^T(D)$ is equivalently modified as $\mbox{\boldmath $\zeta$}_k=\mbox{\boldmath $\tilde e$}_k\tilde H^T(D)$. Note that the correspondence between $\{\mbox{\boldmath $e$}_k\}$ and $\{\mbox{\boldmath $\tilde e$}_k\}$ is one-to-one ($\{e_k^{(j)}\}$ is cyclically shifted). Hence, the original error-path $\mbox{\boldmath $e$}=\{\mbox{\boldmath $e$}_k\}$ is indirectly represented using the reduced tail-biting error-trellis based on $\tilde H^T(D)$. (Accordingly, the restoration in {\it (iv)} is required.) Notice that the overall constraint length $\tilde \nu^{\perp}$ of $\tilde H(D)$ is not more than $\nu^{\perp}$. Thus we have shown the following.
\begin{pro}
Let $T_{tb}^{(e)}$ be a tail-biting error-trellis based on $H^T(D)$, where the $j$th column of $H(D)$ has a monomial factor $D^{l_j}$. Also, suppose that $\tilde \nu^{\perp}<\nu^{\perp}$. Then $T_{tb}^{(e)}$ can be reduced by cyclically shifting the $j$th subsequence of $\{\mbox{\boldmath $e$}_k\}$ by $l_j$ time units. In this case, the reduced error-trellis $\tilde T_{tb}^{(e)}$ is again tail-biting.
\end{pro}

\subsection{Error-Trellis Reduction Using Backward-Shifted Error-Subsequences}
\begin{figure}[tb]
\begin{center}
\includegraphics[width=8.0cm,clip]{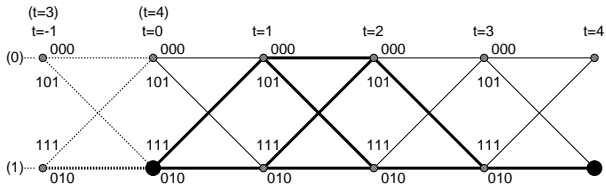}
\end{center}
\caption{Reduced tail-biting code-trellis based on $\tilde G_1(D)$.}
\label{Fig.5}
\end{figure}
A reduced tail-biting error-trellis can be constructed not only using forward-shifted error-subsequences but also using ``backward-shifted'' error-subsequences [9]. For example, consider the parity-check matrix in (8):
$$
H_2(D)=\left(
\begin{array}{ccc}
D^2+D^3& D & 1 \\
D^2& 1+D+D^2 & D^2 
\end{array}
\right) .
$$
Since, the first column has the monomial factor $D^2$, $H_2(D)$ can be reduced to
\begin{equation}
\tilde H_2(D)=\left(
\begin{array}{ccc}
1+D& D & 1 \\
1& 1+D+D^2 & D^2 
\end{array}
\right)
\end{equation}
by dividing the first column of $H_2(D)$ by $D^2$. On the other hand, we can ``multiply'' the second and third columns by $D^2$. Note that this corresponds to backward-shifting by two time units in terms of the original $\{e_k^{(j)}\}~(j=2, 3)$ and we have
\begin{equation}
H_2'(D)=\left(
\begin{array}{ccc}
D^2+D^3& D^3 & D^2 \\
D^2& D^2+D^3+D^4 & D^4 
\end{array}
\right) .
\end{equation}
This matrix can be reduced to an equivalent canonical parity-check matrix $\tilde H_2(D)$. (Note that the first and second rows of $H_2'(D)$ are just delayed versions of the first and second rows of $\tilde H_2(D)$.)

\section{Reduction of Tail-Biting Code-Trellises}
A code-trellis for a tail-biting convolutional code and the corresponding error-trellis are dual to each other. Hence, the reduction of tail-biting code-trellises is also possible. For example, consider the generator matrix $G_1(D)$ in (1). Observe that the first and second columns of $G_1(D)$ have the monomial factor $D$. This fact enables reduction of the original tail-biting code-trellis. Let $u_k$ and $\mbox{\boldmath $y$}_k=(y_k^{(1)}, y_k^{(2)}, y_k^{(3)})$ be the information and code symbol at time $k$, respectively. Then the relation $\mbox{\boldmath $y$}_k=u_kG_1(D)$ is equivalently modified as $\mbox{\boldmath $\tilde y$}_k=u_k\tilde G_1(D)$. Here,
\begin{equation}
\mbox{\boldmath $\tilde y$}_k=(\tilde y_k^{(1)},\tilde y_k^{(2)},y_k^{(3)})\stackrel{\triangle}{=}(y_{k+1}^{(1)},y_{k+1}^{(2)},y_k^{(3)})
\end{equation}
 and $\tilde G_1(D)$ is defined as
\begin{equation}
\tilde G_1(D)=(1+D, D, 1+D) .
\end{equation}
Using a similar argument as that in Section III-A, a reduced tail-biting code-trellis associated with the one in Fig.1 is constructed. The resulting reduced code-trellis is shown in Fig.5, where bold lines correspond to the original code subtrellis with $\mbox{\boldmath $\beta$}_0=\mbox{\boldmath $\beta$}_4=(1,1)$. Note that the first two labels on each branch of the error-path are shifted to the left (i.e., backward-shifted) by the unit time. Accordingly, the path-segment from $t=3$ to $t=4$ is restricted to $010$. We see that this specific example can be directly extended to general cases. We also remark that the reduction of a tail-biting code-trellis and that of the corresponding tail-biting error-trellis can be accomplished simultaneously, if reduction is possible (cf. [10]).

\section{Conclusion}
In this paper, we have discussed the reduction of error-trellises for tail-biting convolutional codes. In the case where a given parity-check matrix $H(D)$ has a monomial factor $D^l$ in some column, we have shown that the associated tail-biting error-trellis can be reduced by cyclically shifting the corresponding error-subsequence by $l$ time units. We have also shown that the obtained reduced error-trellis is again tail-biting. Moreover, we have shown that trellis-reduction is also accomplished using backward-shifted error-subsequences. The proposed method has been applied to concrete examples and it has been confirmed that each subtrellis is successfully embedded in the reduced tail-biting error-trellis. Finally, we have shown that the associated tail-biting code-trellis can equally be reduced using shifted code-subsequences. We remark that the convolutional code specified by a parity-check matrix $H(D)$ with the form discussed in the paper has a relatively poor distance property. We also remark that such parity-check matrices appear, for example, in [13].






%

\end{document}